\documentclass[twocolumn]{aastex631}

\usepackage{graphicx}   
\usepackage{amsmath}    
\usepackage{amssymb}    
\usepackage[T1]{fontenc} 
\usepackage{academicons}
\usepackage{xcolor}
\usepackage{hyperref}
\usepackage{multirow}
\usepackage{float}

\newcommand{\ha}{H$\alpha$}
\newcommand{\hb}{H$\beta$}

\newcommand{\orcid}[1]{\href{https://orcid.org/#1}{\textcolor[HTML]{A6CE39}{\aiOrcid}}}

\def\ltsima{$\buildrel<\over\sim$}
\def\lsim{\lower.5ex\hbox{\ltsima}~}
\def\gtsima{$\buildrel>\over\sim$}
\def\gsim{\lower.5ex\hbox{\gtsima}~}

\def\sfrha{SFR$_{\rm H\alpha}$}
\def\sfruv{SFR$_{\rm UV}$}

\def\msolyr{M$_{\odot}$~yr$^{-1}$}
\def\msol{M$_{\odot}$}
\def\mstar{M$_{\star}$}

\def\teff{\ifmmode T_{\rm eff} \else $T_{\mathrm{eff}}$\fi}

\def\ha{H$\alpha$} 
\def\hb{H$\beta$}

\def\cm2{cm$^{-2}$}

\def\ewo3{$EW_{\mathrm{[O\textsc{iii}]}}$}

\def\oiii{O{\sc iii}}
\def\oii{O{\sc ii}}
\def\nii{N{\sc ii}}

\def\nh{\ifmmode N_{\mathrm{HI}}\else $N_{\mathrm{HI}}$\fi}

\def\vexp{\ifmmode v_{\rm exp} \else v$_{\rm exp}$\fi}
\def\taua{\ifmmode \tau_{a}\else $\tau_{a}$\fi}

\newcommand{\jwst}{{\em JWST}}

\usepackage{newtxtext,newtxmath}
\graphicspath{{./}{figures/}}

\begin{document}

\title{The Extreme Low-mass End of the Mass-Metallicity Relation at $z\sim7$}

\correspondingauthor{Iryna Chemerynska} 
\email{E-mail: iryna.chemerynska@iap.fr}

\author[0009-0009-9795-6167]{Iryna Chemerynska} 
\affiliation{Institut d'Astrophysique de Paris, CNRS, Sorbonne Universit\'e, 98bis Boulevard Arago, 75014, Paris, France}

\author[0000-0002-7570-0824]{Hakim Atek}
\affiliation{Institut d'Astrophysique de Paris, CNRS, Sorbonne Universit\'e, 98bis Boulevard Arago, 75014, Paris, France}

\author[0000-0001-8460-1564]{Pratika Dayal}
\affiliation{Kapteyn Astronomical Institute, University of Groningen, 9700 AV Groningen, The Netherlands}

\author[0000-0001-6278-032X]{Lukas J. Furtak}
\affiliation{Department of Physics, Ben-Gurion University of the Negev, P.O. Box 653, Be’er-Sheva 84105, Israel}

\author[0000-0002-1109-1919]{Robert Feldmann}
\affiliation{Department of Astrophysics, University of Zurich, Zurich, CH-8057, Switzerland}

\author[0000-0002-5612-3427]{Jenny~E.~Greene}
\affiliation{Department of Astrophysical Sciences, Princeton University, 4 Ivy Lane, Princeton, NJ 08544, USA}

\author[0000-0003-0695-4414]{Michael V. Maseda}
\affiliation{Department of Astronomy, University of Wisconsin-Madison, 475 N. Charter St., Madison, WI 53706, USA}

\author[0000-0003-2804-0648 ]{Themiya Nanayakkara}
\affiliation{Centre for Astrophysics and Supercomputing, Swinburne University of Technology, PO Box 218, Hawthorn, VIC 3122, Australia}

\author[0000-0001-5851-6649]{Pascal A. Oesch}
\affiliation{Department of Astronomy, University of Geneva, Chemin Pegasi 51, 1290 Versoix, Switzerland}
\affiliation{Cosmic Dawn Center (DAWN), Copenhagen, Denmark}

\author[0000-0001-7201-5066]{Seiji Fujimoto}
\altaffiliation{Hubble Fellow}
\affiliation{Department of Astronomy, The University of Texas at Austin, Austin, TX, USA}

\author[0000-0002-2057-5376]{Ivo Labb\'e}
\affiliation{Centre for Astrophysics and Supercomputing, Swinburne University of Technology, Melbourne, VIC 3122, Australia}

\author[0000-0001-5063-8254]{Rachel Bezanson}
\affiliation{Department of Physics and Astronomy and PITT PACC, University of Pittsburgh, Pittsburgh, PA 15260, USA}

\author[0000-0003-2680-005X]{Gabriel Brammer}
\affiliation{Cosmic Dawn Center (DAWN), Copenhagen, Denmark}
\affiliation{Niels Bohr Institute, University of Copenhagen, Jagtvej 128, Copenhagen, Denmark}

\author[0000-0002-7031-2865]{Sam E. Cutler}
\affiliation{Department of Astronomy, University of Massachusetts, Amherst, MA 01003, USA}

\author[0000-0001-6755-1315]{Joel Leja}
\affiliation{Department of Astronomy \& Astrophysics, The Pennsylvania State University, University Park, PA 16802, USA}
\affiliation{Institute for Computational \& Data Sciences, The Pennsylvania State University, University Park, PA 16802, USA}
\affiliation{Institute for Gravitation and the Cosmos, The Pennsylvania State University, University Park, PA 16802, USA}

\author[0000-0002-9651-5716]{Richard Pan}
\affiliation{Department of Physics and Astronomy, Tufts University, 574 Boston Ave., Medford, MA 02155, USA}

\author[0000-0002-0108-4176]{Sedona H. Price}
\affiliation{Department of Physics and Astronomy and PITT PACC, University of Pittsburgh, Pittsburgh, PA 15260, USA}

\author[0000-0001-9269-5046]{Bingjie Wang }
\affiliation{Department of Astronomy \& Astrophysics, The Pennsylvania State University, University Park, PA 16802, USA}
\affiliation{Institute for Computational \& Data Sciences, The Pennsylvania State University, University Park, PA 16802, USA}
\affiliation{Institute for Gravitation and the Cosmos, The Pennsylvania State University, University Park, PA 16802, USA}

\author[0000-0003-1614-196X]{John R. Weaver}
\affiliation{Department of Astronomy, University of Massachusetts, Amherst, MA 01003, USA}

\author[0000-0001-7160-3632]{Katherine E. Whitaker}
\affiliation{Department of Astronomy, University of Massachusetts, Amherst, MA 01003, USA}
\affiliation{Cosmic Dawn Center (DAWN), Denmark}

\begin{abstract}

The mass-metallicity relation (MZR) provides crucial insights into the baryon cycle in galaxies and provides strong constraints on galaxy formation models. We use \jwst\ NIRSpec observations from the UNCOVER program to measure the gas-phase metallicity in a sample of eight galaxies during the epoch of reionization at $z=6-8$. Thanks to the strong lensing of the galaxy cluster Abell 2744, we are able to probe extremely low stellar masses between $10^{6}$ and $10^{8}$\msol. Using strong lines diagnostics and the most recent \jwst\ calibrations, we derive extremely-low oxygen abundances ranging from 12+log(O/H) = 6.7 to 7.8. By combining this sample with more massive galaxies at similar redshifts, we derive a best-fit relation of 12+{\rm log(O/H)} = $-0.076_{-0.03}^{+0.03} \times ({\rm log}(M_{\star}))^2+ 1.61_{-0.52}^{+0.52}\times {\rm log}(M_{\star})-0.26_{-0.10}^{+0.10}$, which becomes steeper than determinations at $z \sim 3-6$ towards low-mass galaxies. Our results show a clear redshift evolution in the overall normalization of the relation, galaxies at higher redshift having significantly lower metallicities at a given mass. A comparison with theoretical models provides important constraints on which physical processes, such as metal mixing, star formation or feedback recipes, are important in reproducing the observations. Additionally, these galaxies exhibit star formation rates that are higher by a factor of a few to tens compared to extrapolated relations at similar redshifts or theoretical predictions of main-sequence galaxies, pointing to a recent burst of star formation. All these observations are indicative of highly stochastic star formation and ISM enrichment, expected in these low-mass systems, suggesting that feedback mechanisms in high-$z$ dwarf galaxies might be different from those in place at higher masses.

\end{abstract}

\keywords{Galaxy formation (595), Galaxy evolution (594), High-redshift galaxies (734), Galaxies (573), Reionization (1383), Gravitational lensing (670), Strong gravitational lensing (1643)}

\section{Introduction} \label{sec:intro}

The chemical composition of the interstellar medium (ISM) is a crucial ingredient in the baryonic cycle within galaxies. A galaxy's metal content depends on some environmental factors, including the acquisition of metals and gas gained from mergers and through inflows from the intergalactic medium (IGM), and the loss of metals and gas through outflows. Valuable insights into galaxy growth can be gained by studying the connection between metallicity (the ratio of metal mass-to-gas mass) and inherent galaxy properties, such as stellar mass and star formation rate (SFR)\citep{maiolino2019}. Therefore, metallicity is sensitive to various physical processes that drive the baryon cycle in galaxies.

The gas-phase metallicity in galaxies is often measured through the oxygen abundance, represented as 12+log(O/H). The relationship between gas-phase oxygen abundance and stellar mass is known as the Mass-Metallicity relation (MZR), which is one of the most fundamental scaling relations.  It underscores the intricate interplay between star formation, gas inflow and outflow, and the overall chemical evolution of galaxies \citep{tremonti04,kewley08}. Galaxy metallicity exhibits a tight correlation with stellar mass, while its scatter is often linked to the star formation rate \citep[e.g.][]{ellison2008, Mannucci2010} and gas mass \citep[e.g.][]{Bothwell13}. As such, it offers crucial insights for theoretical studies of galaxy formation, which need to balance these processes to reproduce the observed properties of galaxies over cosmic history \citep{Lilly2013,somerville15,ma16,ucci23}. Furthermore, it has been suggested that the MZR is a two-dimensional representation of a deeper three-dimensional relationship that connects stellar mass, gas-phase metallicity, and the instantaneous SFR, known as the fundamental metallicity relation (FMR)\citep[e.g.][]{ellison2008,lara-lopez2010,Dayal10, Mannucci2010,Hunt2012, Hunt2016,Curti2020}. 
However, characterizing these scaling relations at high redshift has been notably restricted to relatively massive galaxies \citep{tremonti04,henry21,sanders21,nakajima23,curti24}. Consequently, it is unclear whether the MZR extends with the same slope to low-mass galaxies, or whether their different star formation histories lead to different baryon cycle and chemical enrichment.

Metallicity measurements ideally require key optical emission lines such as [\oiii]$\lambda\lambda$5007,4959, [\oiii]$\lambda$4363, [\oii]$\lambda \lambda$3726,37291, and the Balmer lines. The \jwst\ can detect auroral lines, such as [\oiii]$\lambda$4363, which are generated by collisions between particles at higher energy levels than those typically observed in galaxy spectra. This line is particularly important for gas-phase metallicity studies based on electron temperature (T$_e$), and the method of determining electron temperatures/metallicities using this line is known as the "direct T$_e$ method" \citep{Peimbert1967,Bresolin09}. 

Detecting the auroral line at high redshifts is challenging, so large galaxy samples usually determine metallicities using strong line diagnostics based on optical nebular lines. These diagnostics are calibrated against metallicities derived using the direct method \citep{Curti17, Curti2020, Sanders20, nakajima22, Laseter24}. Also, the metallicity calibrations are expected to evolve with redshift. There is now substantial evidence suggesting that, at a fixed metallicity, the ionization conditions of the ISM are evolving to a more extreme state at around z=2  \citep{Steidel14, Shapley15,Sanders16,strom17,Sanders20}. In particular, star-forming galaxies at $ z \sim 2-3$ have higher N/O at fixed excitation than $z \sim 0$ galaxies with similar ionizing spectra. This implies the presence of more intense ionizing radiation fields at fixed N/O and O/H levels compared to typical local galaxies. The results of \citet{Steidel14}, suggest that the systematic offset of high-z galaxies relative to star-forming galaxies in the low-redshift universe in the N2-BPT ([\oiii]/\hb versus the [\nii]/\ha) plane cautions against using the common strong-line metallicity relations for high-redshift galaxies, since the calibrations are designed to reproduce the local N2-BPT sequence \citep{strom17}. To address this issue, calibrations based on electron temperature $T_e$ were developed using low-redshift galaxies that exhibit extreme line ratios or have similar SFR properties to those of typical high-redshift objects  \citep{Bian18, nakajima22}. 
Recent \jwst\ studies have provided important steps towards such calibrations by observing the ratio of [\oiii]$\lambda$4363 to the stronger, lower energy level lines of [\oiii]$\lambda \lambda$4959,5007 in $z=2-8$ galaxies \citep{sanders24} and $z=9.5$ \citep{Laseter24}. This has allowed in particular to investigate the mass-metallicity relation at $z>6$. For example, \citet{nakajima23} and \citet{curti23} used a large sample of galaxies from various \jwst\ programs (ERO, CEERS, GLASS, JADES) to explore the MZR at these redshifts. Additionally, \citet{nakajima22, Langeroodi23, heintz23,fujimoto23} found that the galaxies at $z > 6$ fall below the mass metallicity relations at $z = 0-3$, which may indicate the possible redshift evolution of the MZR. However, all these studies are restricted to relatively massive galaxies and fail to explore the low-mass regime. Extending the MZR to dwarf galaxies ($M_{\star} < 10^{8} \ M_{\odot}$) provides important leverage for constraining the MZR slope, as the effects of star formation feedback are expected to be more pronounced in these low-mass galaxies due to their weaker gravitational potential \citep{ucci23}. 

Here, we explore for the first time the mass-metallicity relation of extremely low-mass galaxies, down to \mstar\ $\sim 10^{6}$ \msol, during the epoch of reionization. These sources benefit from the strong gravitational magnification of the galaxy cluster Abell 2744 and deep NIRSpec spectroscopic observations from the \jwst\ UNCOVER survey \citep{bezanson22}.

The paper is organized as follows: in Section \ref{sec:Observations} we describe the imaging data set used in the study and the lensing model is covered in Section \ref{sec:Gravitational lensing}. In Section \ref{sec:Metalicity}, we utilize strong emission line ratios to derive metallicities for the \jwst\ objects with improved NIRSpec spectra. Then we examine the MZR and its correlation with sSFR. In Section \ref{sec:sfr-m}, we analyze the SFR-Mass relations alongside the \sfrha/\sfruv\ ratio. The implications are discussed in Section \ref{sec:implications}. The conclusion is given in Section \ref{sec:summary}.

Throughout this work, we assume a flat $\Lambda$CDM cosmology with $H_0$ = 70 km s$^{-1}$ Mpc$^{-1}$, $\Omega_{M}$ = 0.3 and $\Omega_{\Lambda}$ = 0.7.

\section{Observations}
\label{sec:Observations}
The UNCOVER dataset contains multi-wavelength NIRCam imaging of the lensing cluster Abell 2744 (A2744) in 6 broadband filters (F115W, F150W, F200W, F277W, F356W, and F444W), one medium-band filter (F410M), and parallel observations with the Near Infrared Imager and Slitless Spectrograph (NIRISS) using five broadband filters (F115W, F150W, F200W, F356W, and F444W). The JWST/NIRSpec low-resolution Prism spectra were collected between July 31st and August 2nd, 2023 as the second phase of the UNCOVER Treasure survey \citep[PIs Labbe \& Bezanson, JWST-GO-2561,][]{bezanson22}. 

All 8 spectroscopic targets were observed with the Micro-Shutter Assembly (MSA), and the MSA observations are separated into 7 paintings, with significant overlap in the center, providing total integration times ranging from 2.7 to 17.4 hours. All sources were assigned three-slitlets, and observations were conducted with a 2-POINT-WITH-NIRCam-SIZE2 dither pattern. The data were analyzed using the JWST/NIRSpec analysis software version 0.6.10 \texttt{msaexp}. The processing was based on level 2 MAST3 products, using the CRDS context file \texttt{jwst\_1100.pmap}. The software performed various basic reduction steps including flat-field, bias, 1/f noise, and snowball correction. It also performed wavelength and photometric calibrations of individual exposure frames \citep{heintz23}.

The photometric component's observational design is detailed in \cite{bezanson22}, the catalogue is explained by \cite{weaver23}, and the photometric redshifts are explored in depth by \cite{wang2023b}. Refer to \cite{Senoda24} for the spectroscopic experimental design and reductions. The original NIRSpec sample has been selected in HFF studies \cite{atek18, bouwens22c}, based on HST observation, in addition to three sources selected from the UNCOVER imaging data based on their photometric redshifts.

\begin{deluxetable*}{lcccccccc}
\tablecaption{The photometric and spectroscopic characteristics of the sample of high-redshift candidates identified through the Abell 2744 cluster. More details about the physical properties are given in \citet{Atek23spec}. The oxygen abundance is derived using the calibration of \citet{sanders24}.\label{tab:props}
}
\tablehead{ \colhead{ID} & \colhead{$M_{\mathrm{UV}}$} & \colhead{$z_{\mathrm {spec}}$}&\colhead{log(\mstar/\msol)}&\colhead{SFR$_{{\rm H}\alpha}$}&\colhead{SFR$_{\rm UV}$}&\colhead{12+log(O/H)}&\colhead{EW (\hb)}\\
\colhead{}&\colhead{AB}&\colhead{}&\colhead{}&\colhead{\msolyr}&\colhead{\msolyr}&\colhead{}&\colhead{\AA}}
\startdata
18924 & $-15.47\pm0.08$ & 7.70 & $5.88_{-0.08}^{+0.13}$ & $0.33\pm0.02$  & $0.01_{-0.07}^{+0.14}$& $6.95\pm0.15$ & $214.8\pm39.9$ \\ 
16155 & $-16.29\pm0.08$ & 6.87 &$6.61_{-0.06}^{+0.07}$ & $0.92\pm0.04$  & $0.04_{-0.06}^{+0.08}$ & $7.01\pm0.19$ & $211.9\pm15.9 $\\ 
23920 & $-16.18\pm0.10$ & 6.00 &$6.30_{-0.03}^{+0.03}$ & $1.32\pm0.04$  & $0.02_{-0.03}^{+0.03}$ & $6.84\pm0.06$ & $334.2\pm28.2 $\\ 
12899 & $-15.34\pm0.11$ & 6.88 &$6.54_{-0.19}^{+0.14}$ & $0.49\pm0.02$  & $0.04_{-0.15}^{+0.12}$ & $6.70\pm0.15$ & $104.4\pm31.8 $ \\ 
 8613 & $-16.97\pm0.04$ & 6.38 &$7.12_{-0.08}^{+0.07}$ & $0.78\pm0.07$  & $0.16_{-0.07}^{+0.08}$ & $6.97\pm0.18$ & $109.0\pm19.7 $\\ 
23619 & $-16.55\pm0.16$ & 6.72 &$6.57_{-0.06}^{+0.10}$ & $0.85\pm0.07$  & $0.04_{-0.05}^{+0.11}$ & $7.19\pm0.20$ & $ 35.2\pm11.7 $\\ 
38335 & $-16.89\pm0.13$ & 6.23 &$6.83_{-0.20}^{+0.25}$ & $1.00\pm0.16$  & $0.07_{-0.15}^{+0.34}$ & $7.46\pm0.32$ & $ 50.3\pm35.5 $ \\ 
27335 & $-17.17\pm0.08$ & 6.75 &$6.73_{-0.08}^{+0.15}$ & $0.73\pm0.10$  & $0.05_{-0.07}^{+0.17}$ & $6.99\pm0.18$ & $ 35.7\pm12.6 $\\ 
\enddata    
\end{deluxetable*}

\section{Gravitational lensing}
\label{sec:Gravitational lensing}
In this work, we adopt the \texttt{v1.1} UNCOVER strong lensing (SL) model of A2744, presented in \cite{furtak23b}, which is publicly available on the UNCOVER website\footnote{\url{https://jwst-uncover.github.io/DR2.html\#LensingMaps}}. The model is based on the parametric approach by \cite{zitrin2015}, which has been updated to be fully analytic and thus not dependent on a fixed grid, which allows for faster computation and with a higher resolution \citep{pascale2022,furtak23b}. The model for A2744 comprises five smooth cluster-scale dark matter halos, centered on the five brightest cluster galaxies (BCGs). It consists of 421 cluster member galaxies identified in the $\sim$ 45 arcmin$^2$ UNCOVER field-of-view, as detailed in \citep{furtak23b}. The \texttt{v1.1} of the model used here is constrained by a total of 141 multiple images (belonging to 48 sources), of which 96 have spectroscopic redshifts \citep{bergamini23a,bergamini23b,roberts-borsani23} and the remaining ones are photometric systems discovered with the UNCOVER imaging \citep{furtak23b,furtak23c}. With these constraints, the model achieves a lens plane image reproduction RMS of $\Delta_{\mathrm{RMS}}=0.51\arcsec$. 

\begin{figure*}
    \centering
    \includegraphics[width=16cm]{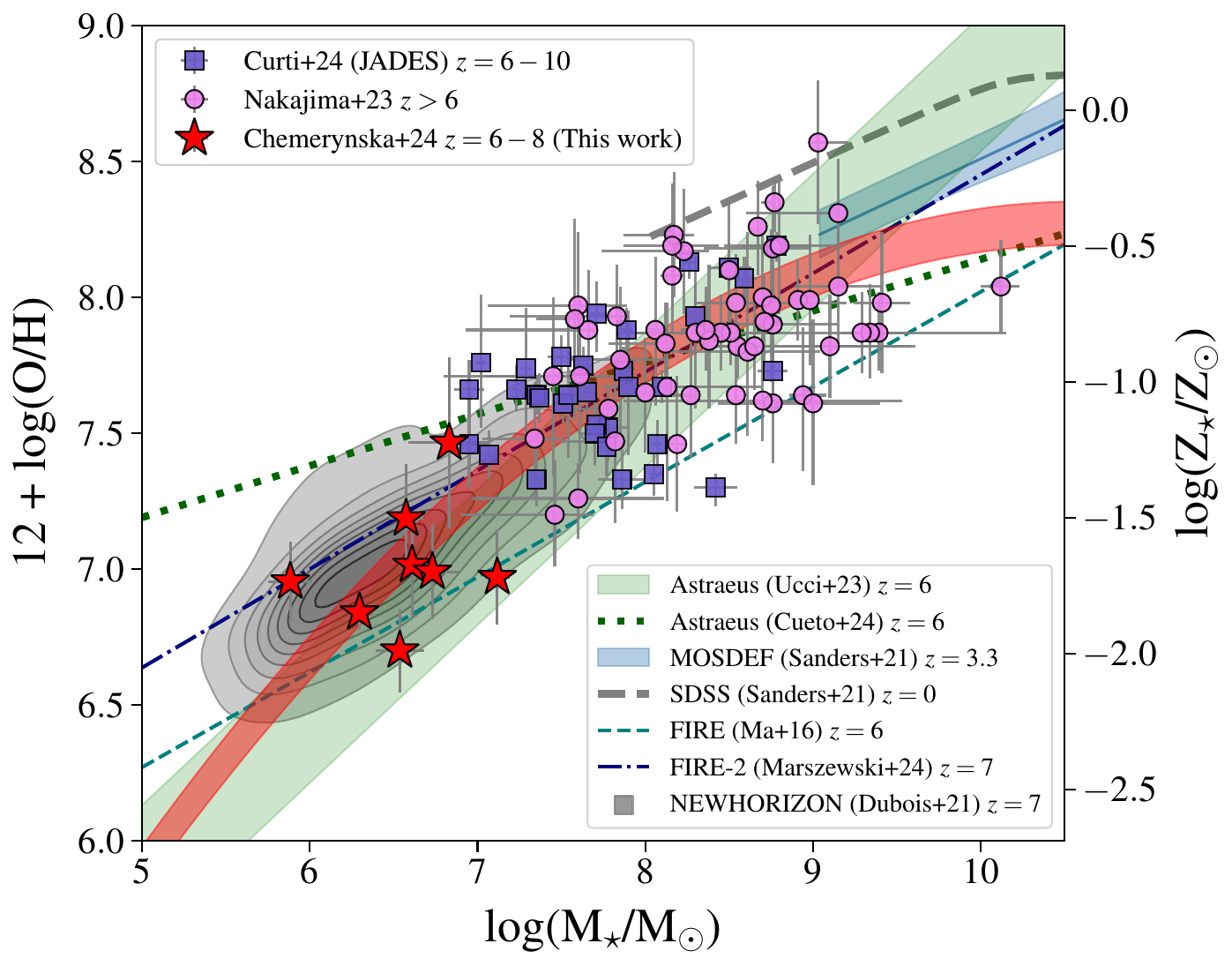}
    \caption{Extending the mass-metallicity relation at $z=6-8$ to the lowest-mass galaxies. The red stars represent measurements of the present sample (best-fit relation is the red-shaded region based on \citet{sanders24} calibration compared to literature results at similar redshifts: the JADES survey \citet[][blue squares]{curti24}, and \jwst\ public surveys \citet[][violet circles]{nakajima23}. We also show for reference lower-redshift determination at $z=0$ from the SDSS \citet[][grey dashed line]{sanders21} and at $z=3.3$ from MOSDEF \citep[][blue-shaded region and line]{sanders21}. A comparison to theoretical predictions of the mass-metallicity relation is also provided: the {\sc FIRE} simulations over the redshift range $z=6$ \citep[][teal dashed line]{ma16}, the {\sc FIRE-2} simulations at redshift 7 \citep[][purple dot-dashed line]{Marszewski24}, the  {\sc Astraeus} determination at $z=6$ \citet[][green-shaded region]{ucci23} assuming SFR $=[0.1-0.5]$ \msolyr\ , the {\sc Astraeus} at $z=6$ assuming an evolving IMF \citep[][green dotted line]{Cueto24} and the {\sc NewHorizon} simulation at the redshift z=7 \citep[][grey-shaded region]{Dubois2021}.
    }
    \label{fig:mz}
\end{figure*}

\section{The Mass-Metallicity relation}
\label{sec:Metalicity}

The main goal of the present paper is to investigate the mass-metallicity relation \citep{tremonti04, Mannucci2010, Perez-montero2013, Lian2015,maiolino2019, Curti2020,baker-maiolino2023} in extremely low-mass galaxies during the epoch of reionization. 
The complete process of the sample selection is as follows. Galaxies are drawn from the Hubble Frontier Fields observations of the A2744 cluster. They have been identified as $z>6$ dropout galaxies using the color-color selection based on the Lyman break technique. This selection was also combined with a photometric redshift estimate that confirmed all these galaxies to be at $z_{best}>6$. On top of that, we have selected only galaxies with a high-magnification factor, typically higher than $\mu=2$ to prioritize intrinsically-faint galaxies. We ended up with a sample of 14 galaxies. Among these, we were able to fit 10 galaxies in the MSA design. We stress that this selection is solely based on the position of the sources in the sky. Sources were chosen to maximize the total number of sources that can be observed in a single MSA mask. This is because we also had other target categories including AGNs, multiple-image systems and very high-redshift galaxies. Among these 10 galaxies that were observed with the NIRSpec MSA, two sources had no usable data (object trace falling in a bad region of the detector or affected by an MSA electrical short). All of the remaining 8 sources were spectroscopically confirmed at redshifts between $z=6$ and $z=8$. Overall, the only notable selection effect introduced by this procedure is favouring extremely faint galaxies in the rest-frame UV. Therefore, we have selected low-luminosity galaxies, that appear to also have very low stellar masses, and we measured their metallicities to probe the extreme low-mass end of the mass-metallicity relation at $z=6-8$.

The stellar mass of each galaxy is derived from spectral energy distribution (SED) fitting using the {\tt Bagpipes} software \citep{Carnall2018,Carnall2019:VANDELS}. The procedure fits simultaneously the spectra and the photometric data points. We fit a polynomial function of order 2 to scale the continuum normalization to the photometry. The model library includes \cite{bc03} stellar population models, the MILES spectral library \citep{Sanchez-Blazquez2006,Falcon-Barroso2011}, CLOUDY nebular emission models \citep{Ferland:2017}, and the \cite{Charlot2000} dust model. We adopt a delayed-$\tau$ star formation history (SFR $\propto^{-t/\tau}$) with the age ($-3<$log(age)$<0.48$) and $\tau$ (0.01 $<\tau<$ 5) as free parameters. By analyzing SED-fitting, we identified that the possible stellar population of our galaxies is very young stellar populations, with stellar ages mostly around a few million years ($t_{50}=1.08-28.66$). This determination is further supported by the blue UV continuum slopes, which range from $\beta = [-2.07, -2.53]$ and are typically associated with young massive stellar populations and low dust attenuation.
A detailed description of the procedure is given in \citet{Atek23spec}. The stellar mass values are shown in Table \ref{tab:props}.  

Regarding the metallicity measurements, we rely on the strong optical lines diagnostics, which have been recently revisited at redshifts greater than $z=6$ \citep[e.g.][]{nakajima23}. For our sources, we mainly detect the following rest-frame optical lines (S/N$\gtrsim$2): \ha+[\nii], [\oiii]$\lambda\lambda$4960,5008, \hb, H$\gamma$, and [\oii]$\lambda$3727. The [\oiii]$\lambda$4363 emission is robustly detected in one source only, but it is blended with the H$\delta$ line. Through simultaneous spectral fitting to the continuum and the emission lines, we estimated robust spectroscopic redshifts between $z\sim$ 6 and $z\sim$ 7.70.

We determine the gas-phase metallicity in our sources by analyzing the strong optical lines. When the auroral lines are not observed, it is possible to use the nebular emission-line diagnostics to evaluate the metallicities in the galaxy sample. In the following, we will explore some widely adopted strong-line diagnostics:
\begin{equation*}
    \begin{array}{l}
		R3 = \log \left( \frac{[\mathrm{OIII}]\lambda5007}{\mathrm{H}\beta} \right)\\
		R2 = \log \left( \frac{[\mathrm{OII}]\lambda3727,3729}{\mathrm{H}\beta} \right)\\
		O32 = \log \left( \frac{[\mathrm{OIII}]\lambda5007}{[\mathrm{OII}]\lambda3727,3729} \right)\\
    \end{array}
\end{equation*}

In order to assess the uncertainties surrounding these indirect diagnostics, we compare different calibrations based on recent \jwst\ observations at high redshift.

\subsection{Metallicity calibrations}

In the present study, we adopt two main calibrations published in \citet{sanders24} and \citet{nakajima22}, which are both based on the direct $T_e$ metallicity measurements with the [\oiii]$\lambda$4363 line. \cite{sanders24} provided the first high-z strong-line calibrations, which are valid over the low-metallicity range of $12+$ log(O/H) $= 7.0-8.4$ and can be applied to samples of star-forming galaxies at $z=2-9$. Regarding the ISM properties, they examine the ionization properties of $\sim$ 160 galaxies at $z=2-9$ from the CEERS (Cosmic Evolution Early Release Science) program \citep{finkelstein23} by using high-to-low ionization emission line ratios such as [\oiii]$\lambda$5007/[\oii]$\lambda$3727, and suggest that galaxies tend to present harder ionizing spectra at higher redshift.

Accordingly, \cite{Laseter24} compared calibrations for R2, O32, R3, and R23 with their sample and assessed the deviation of each calibration from \cite{sanders24}. They found that the R3 and R23 calibrations from \cite{sanders24} do visually trace the upper envelope of their sample, whereas other local calibrations tend to underestimate these ratios. \cite{Laseter24} claim that the set of calibrations presented by \cite{sanders24}, especially the R3 and R23 diagnostics, are now able to offer a more precise depiction of the distribution of galaxies with direct metallicities in the high-z Universe. It is clear that larger samples of direct metallicity measurements will be needed to obtain a more robust calibration at these redshifts.

We also compare the results with other calibrations derived by \cite{nakajima22}, which are applicable to the low-mass metal-poor galaxies. The metallicities are anchored with the direct-method measurements. The gas metallicity diagnostics were established using a combination of local SDSS galaxies and the largest compilation of extremely metal-poor galaxies (XMPGs) identified by the Subaru EMPRESS survey. By using reliable metallicity measurements from the direct method for low-z galaxies, they derive the relationships between strong optical-line ratios and gas-phase metallicity over the range of 12+log(O/H) = $6.9-8.9$ and explore the mass range of approximately $10^{7.5}-10^{9.5}$ \msol. In addition to the direct method, they rely on the rest-frame equivalent widths (EWs) of \hb\ as an additional parameter to control the ionizing properties of the galaxies.  This is because EW(\hb) is sensitive to the current efficiency of massive-star formation and is well correlated with the ionization state as probed by e.g., O32 \citep[e.g.,][]{Nakajima2014,mingozzi2020,nakajima22}.

Additionally, we calculated the metallicities according to a prescription based on the comprehensive emission-line catalogs of galaxies from the IllustrisTNG simulation. This includes ionization by stars, active galactic nuclei, and shocks to reassess the calibrations of optical metallicity estimators at redshifts $0<z<8$ \citep{Hirschmann23}.
This calibration was also confronted to recent \jwst\ results at $4<z<9$ \citep{sanders24,curti24}. The strong-line diagnostics were estimated on metallicities $7\lesssim 12+$log(O/H) $\lesssim 9$ and agreed well with observational results at metallicities below $12+$log(O/H) $\sim8$.

These additional calibrations and prescriptions are explored in detail in Appendix \ref{app:stong-line}. The impact of adopting different calibrations on the MZR relation is also discussed in Appendix \ref{app:MZR}.

\begin{figure}
    \centering
    \includegraphics[width =\linewidth]{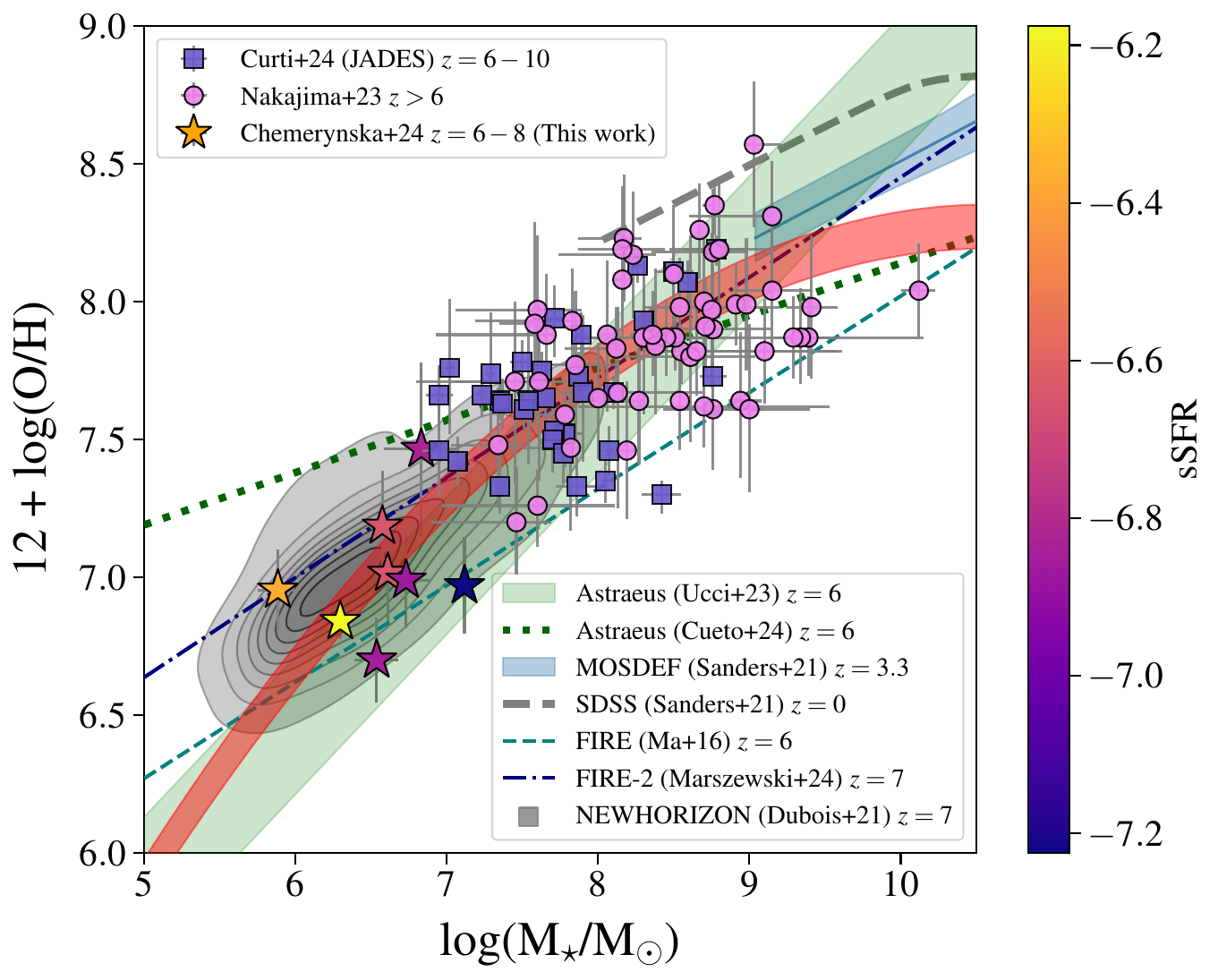}
    \caption{The impact of SFR on the mass-metallicity relation. The stars represent measurements of the present sample, which are colour-coded according to their sSFR. The rest of the legend is identical to Figure \ref{fig:mz}.
    }
    \label{fig:FMR}
\end{figure}

\subsection{Metallicity measurements}

First, we measure the oxygen abundance using the R3 calibration of \citet{sanders24}. For a given value of R3, the calibration defines two metallicity solutions. Although these sources have likely low metallicities, we use the O32 ratio to distinguish between the two branches. For most of the sources, the [\oii] is not detected, which provides a lower limit on O32, which is found to vary in the range O32 $=[0.5-1.8]$. Using the O32 metallicity indicator, the resulting values are all compatible with the low-metallicity branch solution. The metallicity measurements are reported in Table \ref{tab:props}.

With the goal of exploring systematic differences between metallicity calibrations at high redshift, we also used \cite{nakajima22} strong line calibration prescriptions. In addition to the R3-Metallicty relation, we attempt to account for the ionization parameter using the \hb\ equivalent width (large EW  > 200\AA\ , small EW < 200\AA). For [\oiii]$\lambda$4363 emitters, the rest-frame EWs(\hb) can range between 10-600 \AA\ \citep[e.g.,][]{maiolino2019,izotov2021b,Laseter2022,nakajima22}. The fit from \cite{nakajima22} for the large EW is based on the most extreme EW(\hb) objects in their calibration sample. We found that the median value of EW(\hb) for our sample is 136\AA\ with the minimum value $\sim 35$\AA\ and the maximum being $\sim 334$\AA\ (see Table \ref{tab:props}).

Furthermore, we also used the calibration proposed by \citet[][JADES]{Laseter24} to test the results obtained from the calibrations mentioned above. For this purpose, we employed an alternative diagnostic based on a different combination of R3 and R2 with a higher dynamic range, defined as $\hat{R} = 0.47\times R2+0.88\times R3$. As mentioned earlier, the [\oii] line is not detected for the majority of the sources. Thus, we calculated the upper limits of $\hat{R}$. The obtained values are found to lie between the results from \cite{nakajima22}, \cite{Hirschmann23} and \cite{sanders24} calibrations. 

With all calibrations, we find extremely low metallicities, ranging from 12+log(O/H) = 6.70 to 7.76, which corresponds to 1\% to 6\% of the solar metallicity. Such low metallicities often suggest that there is likely strong ionizing radiation from massive stars. Due to their pristine gas conditions, these distant low-mass galaxies are expected to be metal-poor.  Another possible reason is that there was not enough time for the pre-enrichment and many metals were lost due to the outflows. 

Comparing these results to the low-redshift galaxies, XMPGs can show similar properties and comparable metallicities. For example, the Extremely Metal-poor Representatives Explored by the Subaru Survey (EMPRESS) has explored the MZR for XMPGs in the local universe \citep{Kojima20}. Their sample probes a low stellar mass regime (log(\mstar/\msol) = 5-7). However, at a fixed stellar mass, their metallicities (12+log(O/H) = 6.9-8.5) span a larger range compared to the present study. Only two of their extreme galaxies show similar metallicities. Uncertainties in our metallicity measurements can explain part of these differences. For instance, the calibration of \citep{nakajima22} produces slightly higher metallicities, which are more in line with the XPMG values (cf. Appendix \ref{app:MZR}). On the other hand, we note that our sample was essentially selected based on the UV luminosity (cf. Section \ref{sec:Metalicity}), whereas the XMPG sample selection was specifically designed to identify low-metallicity galaxies.   

In the recent study of XMPGs in the Dark Energy Spectroscopic Instrument (DESI) Early Data, \citet{Zou24} analyzed a large sample of galaxies at $z<1$, for which metallicities are measured using the direct method. Again, while a significant number of their galaxies is located below the local MZR for normal galaxies, they still have higher metallicities than our galaxies. Overall, given the different selection methods, our galaxies generally show metallicities comparable to XMPGs in the local Universe, and even to a few extreme cases, which may be the only sources that may serve as local analogues of high-redshift low-mass galaxies.

\subsection{Extending the mass-metallicity to low-mass galaxies at $z\sim7$}

We measured the gas-phase metallicity using the R3 = [\oiii]/\hb\ line ratio based on the most recent calibrations \cite{sanders24,nakajima22}. In Figure \ref{fig:mz} we report the oxygen abundance, in units of 12+log(O/H) as a function of the stellar mass, mass-metallicity relation, together with literature results for more massive galaxies. These measurements are based on the \citet{sanders24} calibration. We included results of more massive galaxies from \citet{curti24} and \citet{nakajima23} at $z > 6$.

\citet{curti24} analysed the gas-phase metallicity properties of a sample of low-stellar-mass (log (\mstar /\msol ) $\lesssim$ 9) galaxies at $3 < z < 10$ observed with JWST/NIRSpec as part of the JADES programme. In order to extend the parameter space, they complement their sample with a sample of $\sim$ 80 high-stellar-mass galaxies at the same redshift from other programmes (\citet[CEERS][]{nakajima23} \citet[ERO][]{curti23} and \citet[GN-z11][]{bunker23b}). In total, the sample consists of 146 galaxies, for which the scaling relations between stellar mass, oxygen abundance, and SFR were explored. The coverage of the low-end of the stellar mass distribution is \mstar /\msol $\approx 10^{6.5}-10^{10}$. In \citet{nakajima23}, they identified 135 galaxies at $z=4-10$ observed by JWST/NIRSpec
. They divided their full sample of ERO, GLASS, and CEERS galaxies into three subsamples at different redshift bins: $z = 4-6, 6-8$, and $8-10$. In each panel, the subsample is further divided into two groups based on their masses and average values. Both studies derived metallicities using calibration based on the local-metal poor galaxies \citep{Curti2020, nakajima22}.

Most theoretical studies predict MZR as a power law \citep[e.g.][]{torrey19}. Analyzing the MZR, derived from observations, over a wide mass range also shows it is well-fitted with higher-order polynomials \citep{tremonti04,kewley08,zahid14,Curti2020}. However, in the mass regime lower than $10^{10}$\msol, many studies adopt a single power law \citep{lee06,sanders21,nakajima22}, as the correlation is roughly linear between $10^{8.5}$ and $10^{10.5}$\msol, and flattening at higher masses. For the first time, we extended MZR to extremely low mass regime. In order to provide a comprehensive picture and account for the possibility of a more complex MZR, we explored second-order relation fit. We also included the single power-law fit in the Appendix \ref{app:MZR}. The second-order fit results in a slightly better $\chi^{2}$=1.35 compared to the single power-law ($\chi^{2}$=1.40). 
The best-fit relation is given by: 
\begin{equation*}
    \begin{array}{lr}
         12+{\rm log(O/H)}=-0.076_{-0.03}^{+0.03} \times ({\rm log}(M_{\star}))^2\\
         + 1.61_{-0.52}^{+0.52}\times {\rm log}(M_{\star})-0.26_{-0.10}^{+0.10}
    \end{array}
\end{equation*}
and is shown with a red shaded region, which represents the $1-\sigma$ uncertainties of the fit. Compared to an extrapolation of the $z = 0$ mass-metallicity relation, these high redshift galaxies clearly present much lower metallicities at a fixed mass, which reflects the redshift-evolution of the MZR and, in turn, the SFR-evolution, also called the fundamental metallicity relation. Recent studies have observed the same trend, where high-z galaxies exhibit lower metallicities than local analogues with equivalent stellar masses of \mstar/\msol$=10^{7.5}-10^{10}$, predicting the relative evolution of the mass-metallicity relation with redshift \citep{heintz23, Langeroodi23}. A comparison to simulations shows that the derived relation is 
matching with the {\sc {\sc Astraeus}} (seminumerical rAdiative tranSfer coupling of galaxy formaTion and Reionization in N-body dark matter simUlationS) predictions at $z=6$ \citep{ucci23} and steeper than the updated {\sc Astraeus} at the low-mass regime, which includes an evolving IMF with redshift and depends on the metallicity of the star-forming gas \citep{Cueto24}. Here we plot the extrapolated relation to lower masses, as the simulation does not go below $10^7$ \msol. Our results are slightly steeper than the {\sc FIRE} (Feedback in Realistic Environment) simulations \citep{ma16}. 
Our results appear to lie between {\sc FIRE} and the recent results of {\sc FIRE-2} simulations \citep{Marszewski24}, which have provided high-quality ISM metallicity prescriptions and enabled the characterization of the MZR. Similarly, our results are in line with the {\sc NewHorizon} simulations \citep{Dubois2021} at similar redshifts. The most recent simulations seem to reproduce better the metal content of the lowest-mass galaxies. They track metal enrichment through the stellar winds, SN Type {\sc II} (SN{\sc II}), SN Type {\sc I}a (SN{\sc I}a) explosions, and asymptotic giant branch (AGB) stars.

A comparison to results at lower redshifts shows a redshift evolution in the overall normalization of the MZR. At a given stellar mass, galaxies at higher redshifts have lower metallicities. Specifically, local galaxies (SDSS) exhibit metallicities approximately 0.5 dex higher than high redshift galaxies.

We note that our metallicity estimates are based on the calibration of strong line diagnostics. These diagnostics depend on the ionization parameter \citep[e.g.,][]{pilyugin16}. Therefore, we expect that any variation in this parameter may introduce variability in the inferred metallicity.

We have also explored the effect of adopting different high$-z$ metallicity calibrations on the MZR best-fit relation. The results based on the prescriptions of \cite{Laseter24} or \cite{nakajima22} are shown in Appendix \ref{app:MZR}. To guide the eye, we also plot a single-power law best-fit relation to the data (using binned data from the literature), to highlight how the metallicity data points move upward and downward.

In addition to the mass-metallicity correlation, a second dependency is observed with the star formation rate. This redshift-invariant fundamental metallicity relation can describe the general evolution of the MZR and the SFR correlation. Such dependence was observed in both observational results \citep{Mannucci2010} and simulations \citep{Garcia24}. In Figure \ref{fig:FMR}, we color-coded our sample by sSFR to identify the secondary dependence. Our galaxies tend to have higher sSFR than the Main Sequence of galaxies. As discussed by \cite{Laseter24}, it is expected that with decreasing metallicities and/or masses, there will be an increase in sSFR. This relationship with sSFR at $z>5$ was also reported in simulations \citep{ucci23}. However, recent \jwst\ observations have challenged this picture at high-redshift, showing deviations from the FMR at $z>3$ \citep[e.g.][]{curti23,heintz23,Morishita24}. In this study, we find weak evidence for the existence of the FMR at $z\sim7$. 

\section{The SFR-$M_{\star}$ relation}
\label{sec:sfr-m}

\begin{figure*}
\centering
\includegraphics[width=0.49\linewidth]{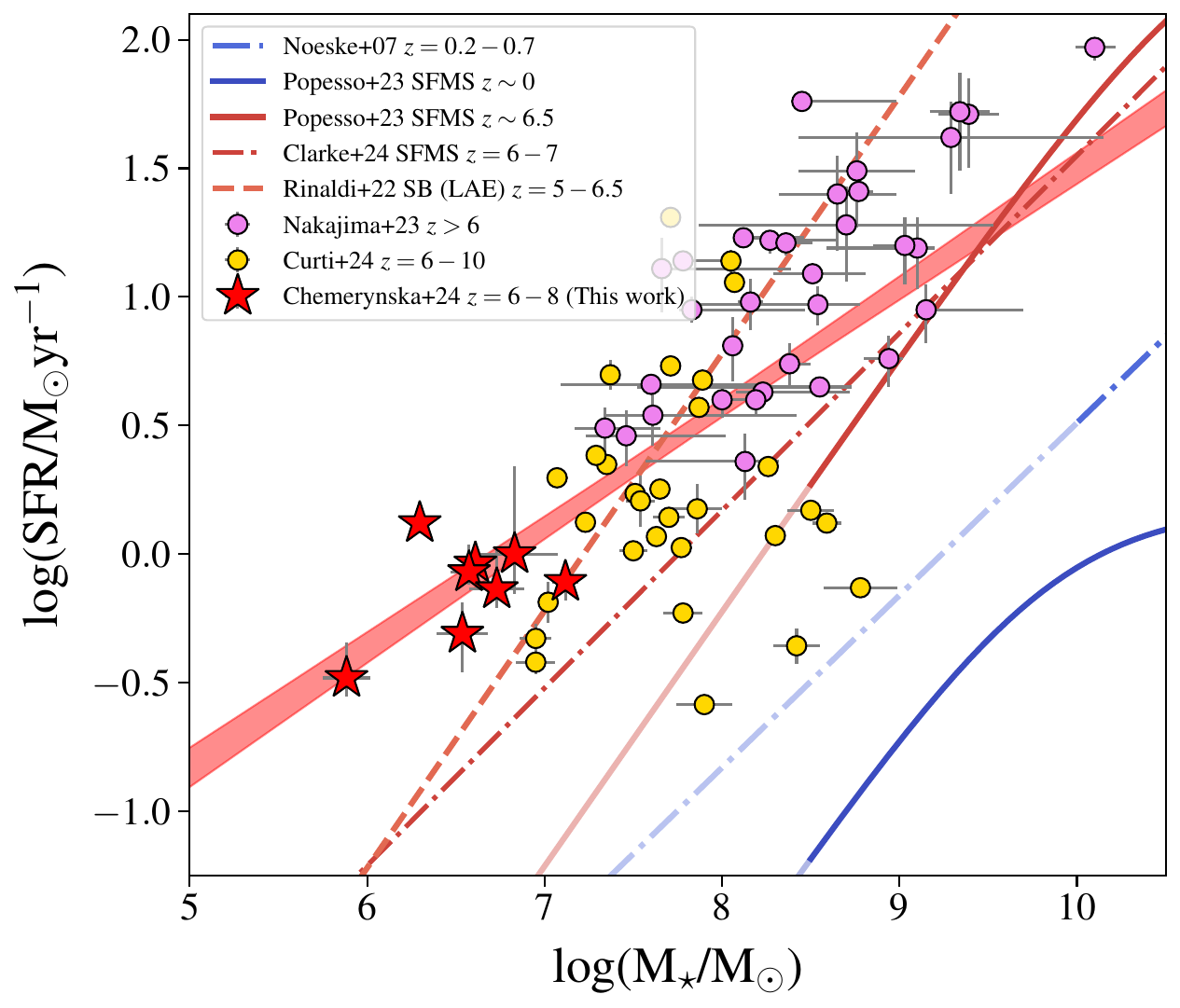}
\includegraphics[width=0.49\linewidth]{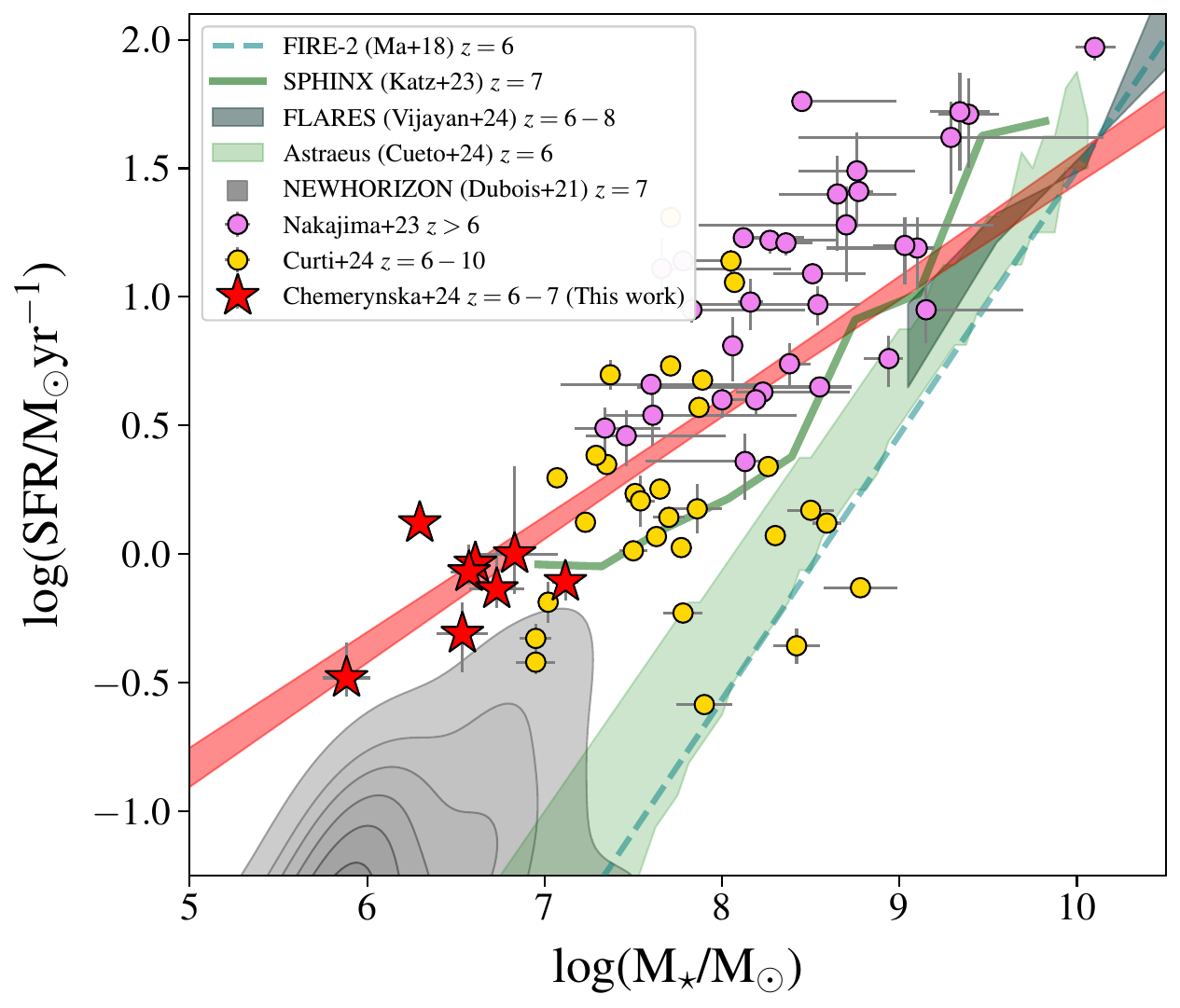}
\caption{The relationship between stellar mass (\mstar) and SFR for our \jwst sample (red stars), including the literature sample compiled from \citet[][violet dots; CEERS]{nakajima23}, and \citet[][gold dots; EROs]{curti24}. {\bf Left:} Comparison between our best-fit galaxy and the Main Sequence of Star-forming galaxies determined from the literature: at $z\sim 0$, and $z\sim 6.5$ \citep[][solid blue and dark red lines, respectively]{popesso2023}, at $6 < z \lesssim 7$ \citep[][dark red dot-dashed line]{Clarke24}, at $0.2<z<0.7$ \citep[][dot-dashed blue line]{Noeske2007}, and starburst cloud at high redshift \citep[][dashed red line]{rinaldi22}. The extrapolation for lower masses is depicted in lighter shades. Literature results are color-coded depending on the redshift. {\bf Right:} We also provided recent results from the {\sc SPHINX} simulation for SFMS for 10Myr at $z=7$ \citep[][green solid line]{Katz23}, the {\sc FLARES} simulation at $z = 6-8$ \citep[][grey shaded region]{vijayan24}\ with SFR timescale 100 Myr, the {\sc Astraeus} at $z=6$ with an evolving IMF \citep[][green-shaded region]{Cueto24}, the {\sc FIRE-2} simulations at the redshift $z=6$ with SFR averaged over 10 Myr \citep[][teal dashed line]{ma18} and the {\sc NewHorizon} simulation for SFR for 10 Myr at the same redshift \citep[][grey-shaded region]{Dubois2021}. The red shaded area on both figures indicates uncertainties in the fitting.}
\label{fig:sfr-m}
\end{figure*}

A strong correlation between the star formation rate and the stellar mass of star-forming galaxies has been established across a wide range of redshift \citep{Brinchmann2004, Noeske2007, whitaker2014,atek22}. The so-called star-forming ``main sequence'' (SFMS) reflects the steady stellar mass buildup in galaxies over hundreds of Myrs. The slope and the dispersion of the relation are good indicators of the star-formation histories in a given population of galaxies. Reproducing the SFR-\mstar\ relation is also an important requirement for galaxy formation models \citep[e.g.,][]{sparre15,Katz23}. The advent of \jwst\ has also allowed us to investigate the existence of this relation out to the highest redshifts \citep{Clarke24}. The present sample extends the high-redshift constraints on the SFMS to extremely low-mass galaxies, allowing us to determine whether dwarf galaxies at $z=6-8$ follow the same SFH as their massive counterparts.   

In Figure \ref{fig:sfr-m}, we plot the correlation between the SFR and \mstar\ for our sample of galaxies (red stars). We compute the SFR using the \ha\ recombination line, while the stellar mass is derived from SED fitting (cf. Section \ref{sec:Metalicity}). We compare our results to the most recent measurements based on \jwst\ observations at similar redshifts \citep{rinaldi22,nakajima23,curti24,Clarke24}. Combining our sample with the literature results of \citet{curti24} and \citet{nakajima23}, which probe higher stellar masses, we derive the following best-fit relation between the SFR and stellar mass:   %

\begin{equation*}
    \log(SFR)=0.47_{-0.06}^{+0.06} \times \log(M\star) - 3.17_{-0.12}^{+0.12}
\end{equation*}
 
For comparison, we also plot the low-redshift parametrization of the SFMS \citep{Noeske2007}, the literature compilation of \citet{popesso2023} both at $z=0$ and $z=6.5$, and finally, the relation derived for starburst galaxies by \cite{rinaldi22} at redshifts 5 to 6.5. The most striking result is the significant offset of this low-mass sample from the literature results and their respective extrapolation to lower masses. At a given stellar mass, the galaxies in the present sample show higher star formation rates, by a factor ranging from a few to tens, compared to SFR-\mstar\ relations at similar redshifts. Our sample is located even above the starburst sample of \citet{rinaldi22}. Perhaps this is not surprising since their sample consists of Ly$\alpha$ emitters (LAEs) for which the SFR is derived from their SED fitting or UV luminosity which traces star formation on different timescales. Additionally, \citet{Rinaldi24} demonstrate that for log(\mstar/\msol) < 7, there is no clear distinction between main sequence (MS) and starburst galaxies. This trend could be explained by an increased burstiness of star formation, which becomes more significant as stellar mass decreases, or by the limitations of current observational depth. When comparing different samples, the same caveat applies to the compilation of \citet{popesso2023} where galaxies at high redshift lack Hydrogen recombination lines for their SFR estimates. The notable exceptions here are the samples of \citet{curti24} and \citet{nakajima23}, for which the SFRs are measured from either \ha\ or \hb\ emission lines and the stellar mass was derived with {\sc BEAGLE} and {\sc Prospector}, respectively.
This is precisely why these samples are included in our fitting of the SFR-\mstar\ relation described above. Besides the offset that may indicate a flattening of this relation at lower masses, the best-fit relation shows a shallower slope than the high-redshift determinations. Because we might expect such offset in galaxies selected by their strong emission lines, it is worth noting that the selection of this sample was only based on the faint UV luminosity (and/or high lensing amplification factors). Therefore, this is clearly an indication of a recent burst of star formation, which is better captured by the \ha\ emission that responds to short-lived massive stars over a few Myrs. When comparing the SFR indicators based on \ha\ and the UV, we observed values ranging between \sfrha/\sfruv\ $\sim5$ and  \sfrha/\sfruv\ $\sim66$. While this might well indicate a bursty-dominated star formation in low-mass galaxies at early times, a larger sample of dwarf galaxies, with stellar masses around $10^{6}$ \msol, is needed to confirm this trend. A statistical sample will also help characterize the duty cycle of this stochastic SFH.  

\section{Comparison with theoretical models}
\label{sec:implications}

Our observed sources have stellar masses ranging between $10^{5.9}$ and $10^{7.1}$ \msol\ with metallicities ranging between 12+log(O/H) = 6.8-7.8. We now compare these physical properties to the predictions of different theoretical models of galaxy formation. We first consider {\sc Astraeus} \citep{hutter21, ucci23}, a semi-numerical model coupling galaxy formation and reionization on 230 Mpc scales which simulates the mass-metallicity relation between $\sim 10^{6.5-10}$ \msol\ in stellar mass tuned against $z\sim 5-10$ observables. The second set of predictions is from the {\sc FIRE} \citep{ma16} zoom-in simulations that can track the mass-metallicity relation between $\sim 10^{3-9}$ \msol\ in stellar mass, which are calibrated against $z\sim 0-3$ data. We also explore the most recent {\sc FIRE-2} suite of simulations \citep{ma18}, tuned for the $z>5$ Universe. Finally, we also plot the mass-metallicity relation for simulated galaxies at $z=6$ from the {\sc NewHorizon} \citep{Dubois2021}, which combines an intermediate volume (16 Mpc)$^3$ and a high-resolution (34 pc) in order to capture the multi-phase structure of the ISM.   

From our metallicity measurements, the bulk of the observed galaxies overlap with the predictions of {\sc Astraeus}. {\sc Astraeus} simulations assume low-mass galaxies to form stars at the maximum limit (at which supernova energy balances the halo binding energy) and perfect, instantaneous metal mixing in the ISM. It incorporates key processes of gas accretion, cooling, star formation, and feedback from supernovae and AGNs. The model does not show a significant effect of reionization feedback at these redshifts. 
In Figure \ref{fig:FMR}, we explored the potential impact of sSFR on the MZR, which is directly related to the FMR. However, with a limited sample size, it is challenging to derive a statistically meaningful 3-parameter relation. Furthermore, we explore the recent implementation of the evolving IMF into {\sc Astraeus} \citep{Cueto24}.  
The IMF evolves in each galaxy according to the metallicity of its star-forming gas and redshift.
It includes dependence on the SN feedback, metal enrichment, and both ionizing and UV radiation. The amount of newly formed metals depends on the quantity of massive stars that explode as SN during the current time step. The simulation is limited to a mass range above $10^7$\msol. Our findings are located below the extrapolation of this simulation. These findings suggest that, in general, the {\sc Astraeus} conforms to a broad range of the MZR relation, where significant variations can be observed at lower masses depending on the adopted model. 

The {\sc FIRE} simulations use a density threshold for star formation at $10-100$ cm$^{-3}$ and allow imperfect mixing of metals in the ISM. On the other hand, the new suite of {\sc FIRE-2} simulations use a density threshold of 1000 cm$^{-3}$ to trigger star formation. This new version tracks the abundances of several metals, which are injected in the ISM via supernovae feedback and stellar winds. In addition, the simulations incorporate subgrid turbulent processes to allow for efficient metal mixing. {\sc FIRE-2} simulations predict metallicities 0.3–0.4 dex higher than {\sc FIRE}. The best-fit MZR relations predicted by the two simulations bracket our measurements around $z=7$.  

The {\sc NewHorizon} cosmological simulations are designed to capture the multi-scale ISM physics in an average-density environment. It includes star formation above a density threshold of 10 cm$^{-3}$ with varying efficiency, which evolves with time to become more bursty at high redshift, as well as feedback from SNe and massive stars. It assumes that SN explodes when a star particle becomes older than 5Myr. Although their statistics are more suited for intermediate-mass galaxies, they also cover the physical properties of galaxies down to $\sim 10^{6}$ \msol. Because the simulations do not track the evolution of individual elements, the oxygen abundance is scaled assuming a solar metallicity. Despite these crude prescriptions, the predictions align remarkably well with our observations.      

Overall, these low-mass galaxies exhibit low gas-phase metallicities, most likely due to low SFRs, which lead to the production of fewer metals in combination with a dilution effect due to gas accretion. In addition, a large fraction of the metals are easily lost due to their shallow potential that enables strong outflows. Theoretical models \citep[e.g.;][]{ucci23}, have shown that the gas mass lost in outflows is higher in low-mass galaxies. If these processes happen on short timescales, then we expect a larger scatter in the MZR at lower masses. It must be noted that none of these models have been tuned to reproduce the unprecedented data presented here for dwarf galaxies at $z\sim 7$. It is therefore heartening to see the reasonable agreement with the data despite their different formalisms for star formation, feedback and metal enrichment. Our observations therefore present a crucial resource to baseline theoretical models.

\section{Summary}
\label{sec:summary}

Combining the strong gravitational lensing of Abell 2744 and ultra-deep NIRSpec observations, we were able for the first time to extend the mass-metallicity relation to extremely low-mass galaxies during the epoch of reionization ($6<z<8$). Our sample consists of 8 galaxies with intrinsic magnitudes between $-17.17<M_{UV}<-15.47$. Using SED fitting of the spectro-photometric data, we derived low stellar masses down to $\sim 10^6$ \msol, corrected for amplification. We measured gas-phase metallicities using strong line diagnostics together with the most recent \jwst\ calibrations \citep{sanders24}. Our measurements yield very low oxygen abundances, in the range 12 + log(O/H) = 6.70 to 7.76, corresponding to 1\% to 6\% of the solar metallicity.

The central goal of the present study is to explore how the mass-metallicity in low-mass galaxies compares to their massive counterparts, in terms of the slope and the normalization. We find a clear offset in the overall normalization of the MZR compared to extrapolations of local or $z=3$ relations based on more massive galaxies, indicating a strong redshift evolution. 

Along these lines, we also investigated the star formation rate and stellar mass relation, called the star formation main sequence. At a given stellar mass, the galaxies in our sample exhibit higher SFR by a factor ranging from a few to tens compared to samples at similar redshifts. This suggests a recent burst of star formation, which reflects short-lived massive stars over a few million years. We also find evidence for bursty star formation by analyzing their $SFR_{H\alpha} /SFR_{UV}$ ratio. However, a larger sample of dwarf galaxies with stellar masses around $10^6$ \msol\ is needed to confirm this trend and characterize the parameters of this stochastic star formation history.

A comparison to galaxy formation models indicates an overall agreement with {\sc NewHorizon} simulations. The {\sc FIRE} and {\sc FIRE-2} suite of simulations which differ in the implementation of several physical processes, including the metal mixing efficiency, encompass most of our measurements, and show a coarse agreement with the slope of the MZR. The median MZR predictions of the {\sc Astraeus} set of simulations, which show a broad range of metallicities based on different models, encompass the observational constraints of the present sample.
 
Overall, these low-mass galaxies exhibit low gas-phase metallicities, likely due to low SFRs that produce fewer metals and potentially episodes of gas accretion. Additionally, a significant fraction of the metals are easily ejected due to strong outflows in low-mass galaxies. 
The reasonable agreement between our data and theoretical models, despite different formalisms for star formation, feedback, and metal enrichment, is encouraging. Our observations thus provide a crucial resource for benchmarking theoretical models.

\begin{acknowledgments}
\noindent This work is based on observations obtained with the NASA/ESA/CSA \textit{JWST} and the NASA/ESA \textit{Hubble Space Telescope} (HST), retrieved from the \texttt{Mikulski Archive for Space Telescopes} (\texttt{MAST}) at the \textit{Space Telescope Science Institute} (STScI). STScI is operated by the Association of Universities for Research in Astronomy, Inc. under NASA contract NAS 5-26555. This work has made use of the \texttt{CANDIDE} Cluster at the \textit{Institut d'Astrophysique de Paris} (IAP), made possible by grants from the PNCG and the region of Île de France through the program DIM-ACAV+. This work was supported by CNES, focused on the \jwst\ mission. This work was supported by the Programme National Cosmology and Galaxies (PNCG) of CNRS/INSU with INP and IN2P3, co-funded by CEA and CNES. IC acknowledges funding support from the Initiative Physique des Infinis (IPI), a research training program of the Idex SUPER at Sorbonne Université. PD acknowledge support from the NWO grant 016.VIDI.189.162 (``ODIN") and warmly thanks the European Commission's and University of Groningen's CO-FUND Rosalind Franklin program.

\end{acknowledgments}

\vspace{5mm}

\section*{Data Availability}
The data underlying this article are publicly available on the \texttt{Mikulski Archive for Space Telescopes}\footnote{\url{https://archive.stsci.edu/}} (\texttt{MAST}), under program ID 2561. The specific observations analyzed can be accessed via \dataset[doi: 10.17909/c9cj-cd41]{https://doi.org/10.17909/c9cj-cd41}.Reduced and calibrated mosaics are also available on the UNCOVER webpage: \url{https://jwst-uncover.github.io/}

\appendix

\section{Strong-line diagnostics}
\label{app:stong-line}

\begin{figure*}
    \centering
    \includegraphics[width=0.49\linewidth]{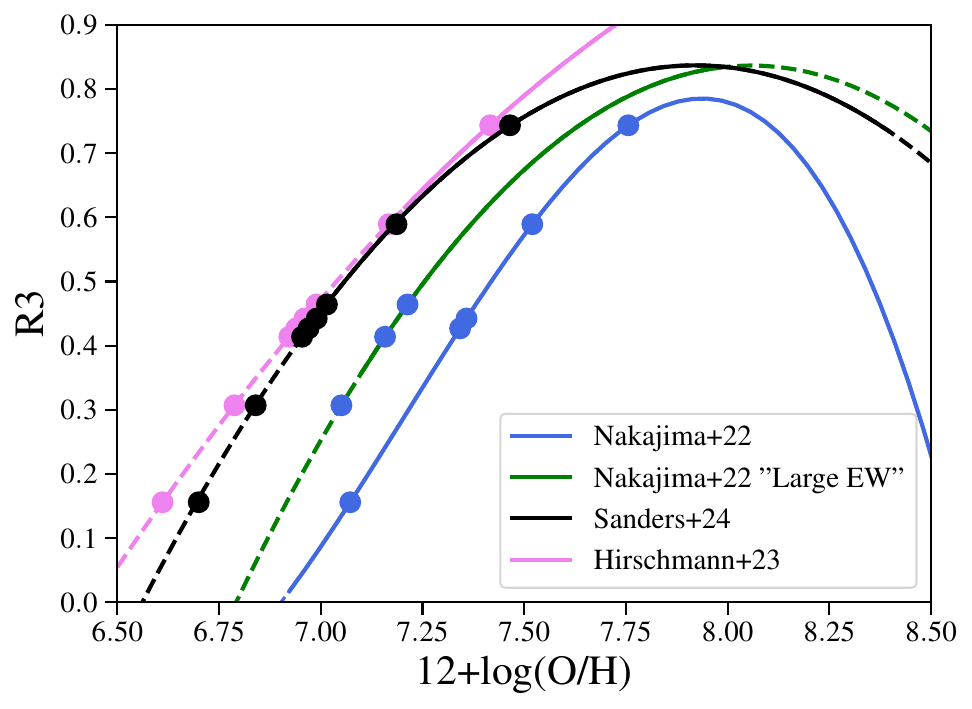}
    \includegraphics[width=0.49\linewidth]{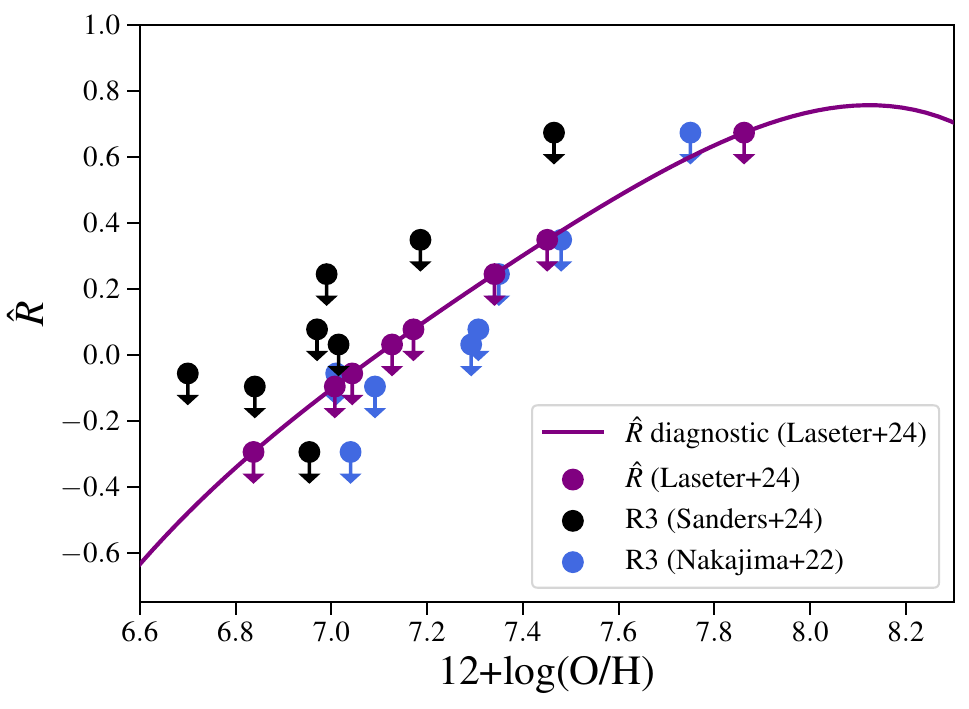}
    \caption{Relationships between metallicity and the strong line rations of R3 and $\hat{R}$ (from left to right).
    }
\label{fig:R3_Rhat}
\end{figure*}

\begin{deluxetable*}{lccccccc}
\tablecaption{Comparison bewteen oxygen abundances, 12+log(O/H), derived using recent high$-z$ calibrations or locally calibrated strong line diagnostics. \label{tab:metallicities}}
\tablehead{ \colhead{ID} & \colhead{\citet{Laseter24}} & \colhead{\citet{nakajima22}} & \colhead{\citet{Hirschmann23}} }
\startdata
18924 & $ 6.84 \pm 0.15 $ & $ 7.16 \pm 0.15 $ & $ 6.92 \pm 0.15 $\\
16155 & $ 7.13 \pm 0.19 $ & $ 7.21 \pm 0.19 $ & $ 6.99 \pm 0.19 $\\
23920 & $ 7.01 \pm 0.06 $ & $ 7.05 \pm 0.06 $ & $ 6.79 \pm 0.06 $\\
12899 & $ 7.04 \pm 0.15 $ & $ 7.07 \pm 0.15 $ & $ 6.61 \pm 0.15 $\\
8613& $ 7.17 \pm 0.18 $ & $ 7.34 \pm 0.18 $ & $ 6.94 \pm 0.18 $\\
23619 & $ 7.45 \pm 0.2 $ & $ 7.52 \pm 0.2 $ & $ 7.17 \pm 0.2 $\\
38355 & $ 7.86 \pm 0.32 $ & $ 7.76 \pm 0.32 $ & $ 7.42 \pm 0.32 $\\
27335 & $ 7.34 \pm 0.18 $ & $ 7.36 \pm 0.18 $ & $ 6.96 \pm 0.18 $\\
\enddata 
\end{deluxetable*}

In section \ref{sec:Metalicity}, we described the high-redshift calibration used in this study. We compared two recent calibrations \citep{sanders24, nakajima22} and simulation results \citep{Hirschmann23} to derive the metallicity of $z=6-8$ galaxies. From those calibrations, we adopted the best fit of the R3 diagnostic (see Figure on the left panel of \ref{fig:R3_Rhat}). For \citet{nakajima22}, we also took into account the dependence on the EW(\hb), as it was described in the study (a blue curve for "All" sample and a green curve for the sample with EW>200\AA). We see that for a fixed value of R3, we obtained different values of metallicity. We adopted the \citet{sanders24} calibration as our fiducial metallicity estimator since the rest of the studies rely on a set of locally calibrated strong line diagnostics, which may overestimate the metallicities. Additionally, we noticed that metallicities predicted with the simulation IllustrisTNG \citep{Hirschmann23} also indicate lower metallicities at high redshifts, which is consistent with the calibration derived by \citet{sanders24}. In order to confirm the metallicities we derived with R3 diagnostic, we also probe the $\hat{R}$, which is a novel calibration derived by \cite{Laseter24} and was earlier introduced by \cite{Curti17} and \cite{Maolino08}.
It is a combination of R2 and R3 in the form:
\begin{equation*}
\hat{R} = \cos(\phi)R2+sin(\phi)R3
\end{equation*}
which is equivalent to a rotation of the R2-R3 plane around the O/H axis. They used the fourth-order polynomial to the resulting $\hat{R}$ ratio versus the metallicity in the form of $\hat{R}=\Sigma_n c_n \cdot x^n$, where $x =12 +\log(O/H)-8.69$ and identify the angle $\phi$ that allows the scatter to be minimized in metallicity from the best fit-relation. In this fit $\phi$ =61.82 deg, which translates into $\hat{R}$ = 0.47R2 + 0.88R3.

As mentioned in Section \ref{sec:Metalicity}, we did not detect the [\oii] line for the majority of our sample, but we were able to derive the upper limits on R2, thus on $\hat{R}$.
In Figure \ref{fig:R3_Rhat} (right-hand panel), we compared the metallicities we obtained using different calibrations. For \citet{sanders24} and \citet{nakajima22} we calculated metallicities using R3 diagnostic and then estimated the $\hat{R}$ by using \citet{Laseter24} fit mentioned above. As $\hat{R}$ provides us with upper limits, it means that true metallicities will have lower values. the metallicities derived from the different calibrations are presented in Table \ref{tab:metallicities}.

\section{Comparison of high-z metallicity calibrations}
\label{app:MZR}
\begin{figure*}[htbp]
\centering
    \includegraphics[width=0.45\linewidth]{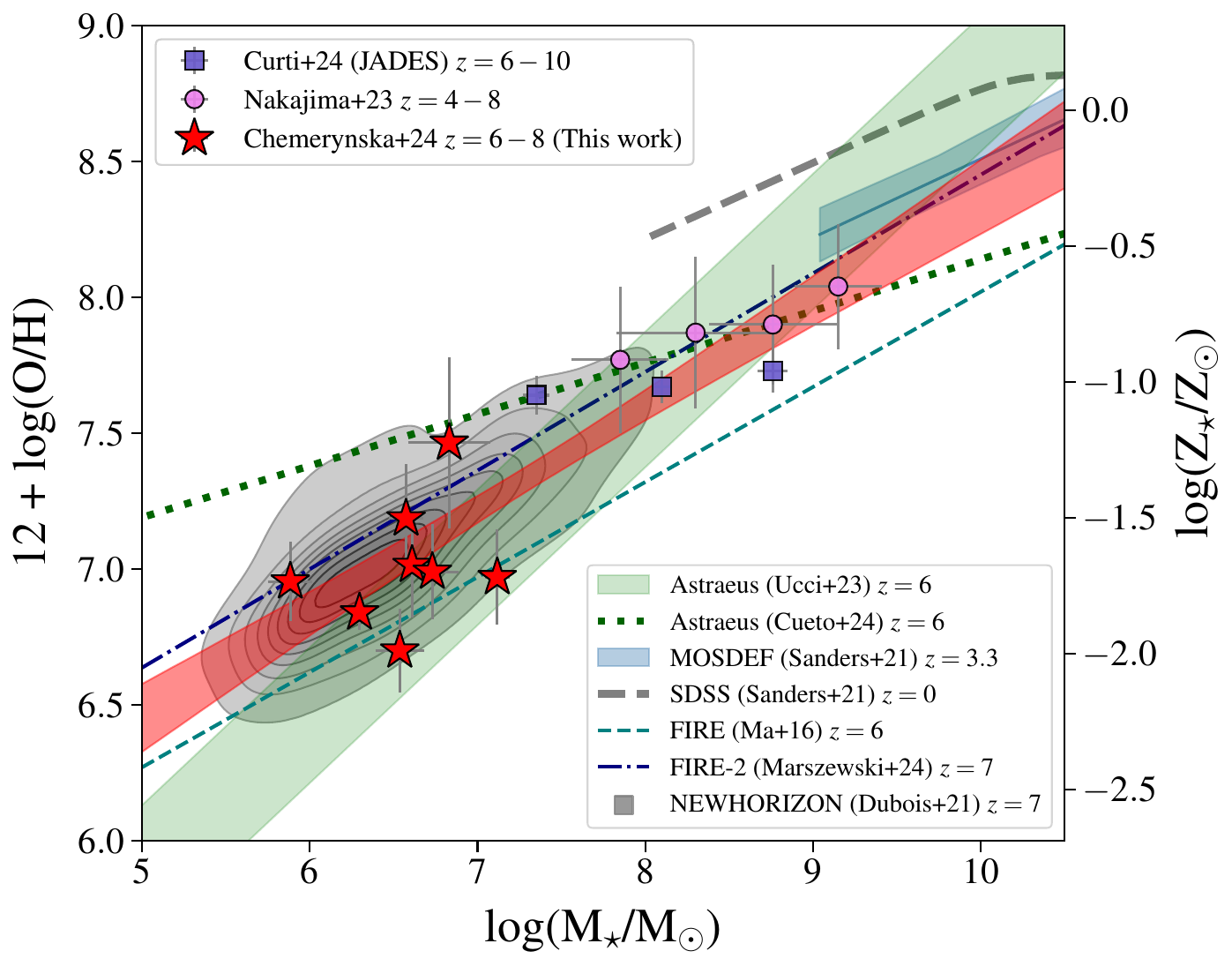}
    \includegraphics[width=0.45\linewidth]{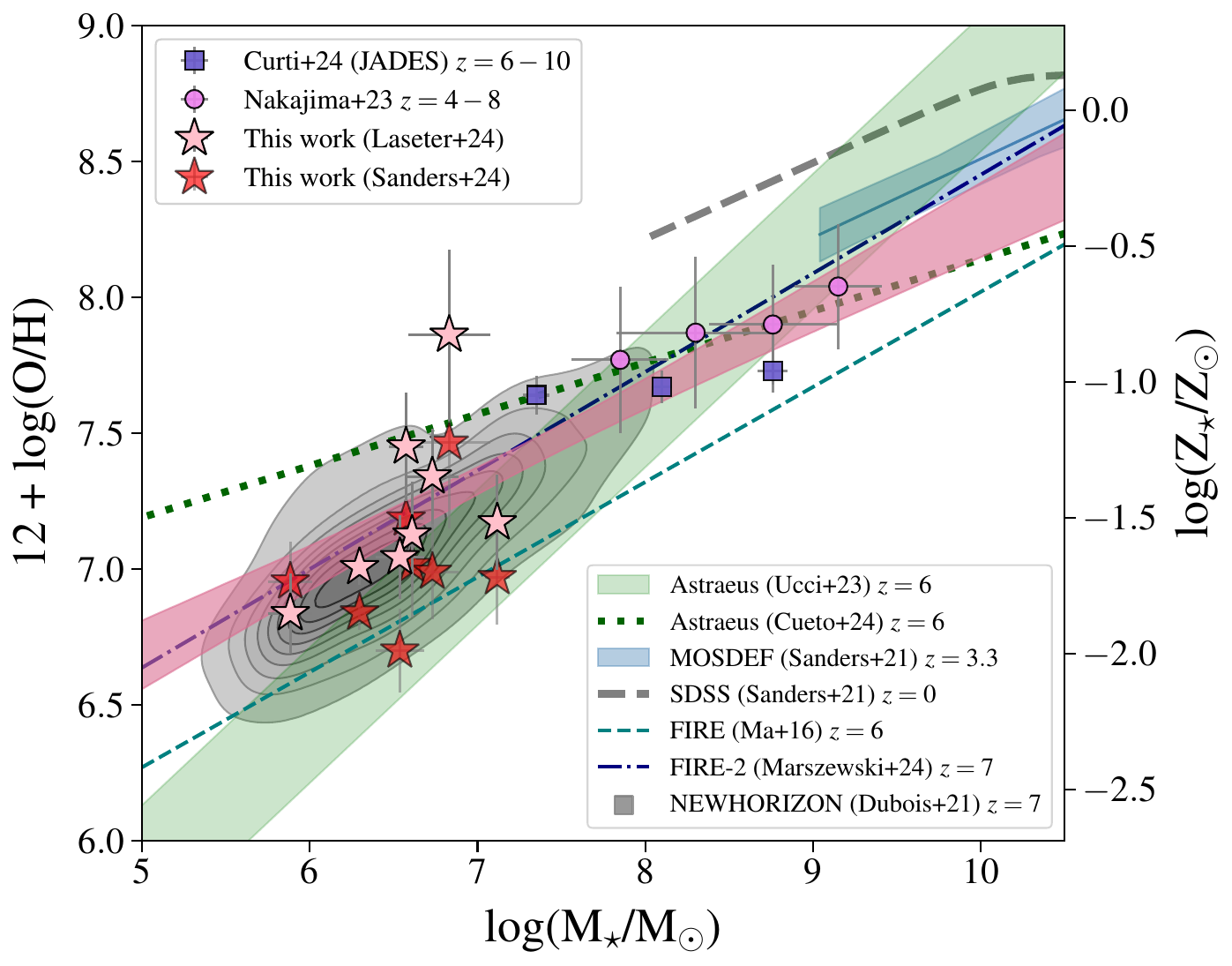}
    \includegraphics[width=0.45\linewidth]{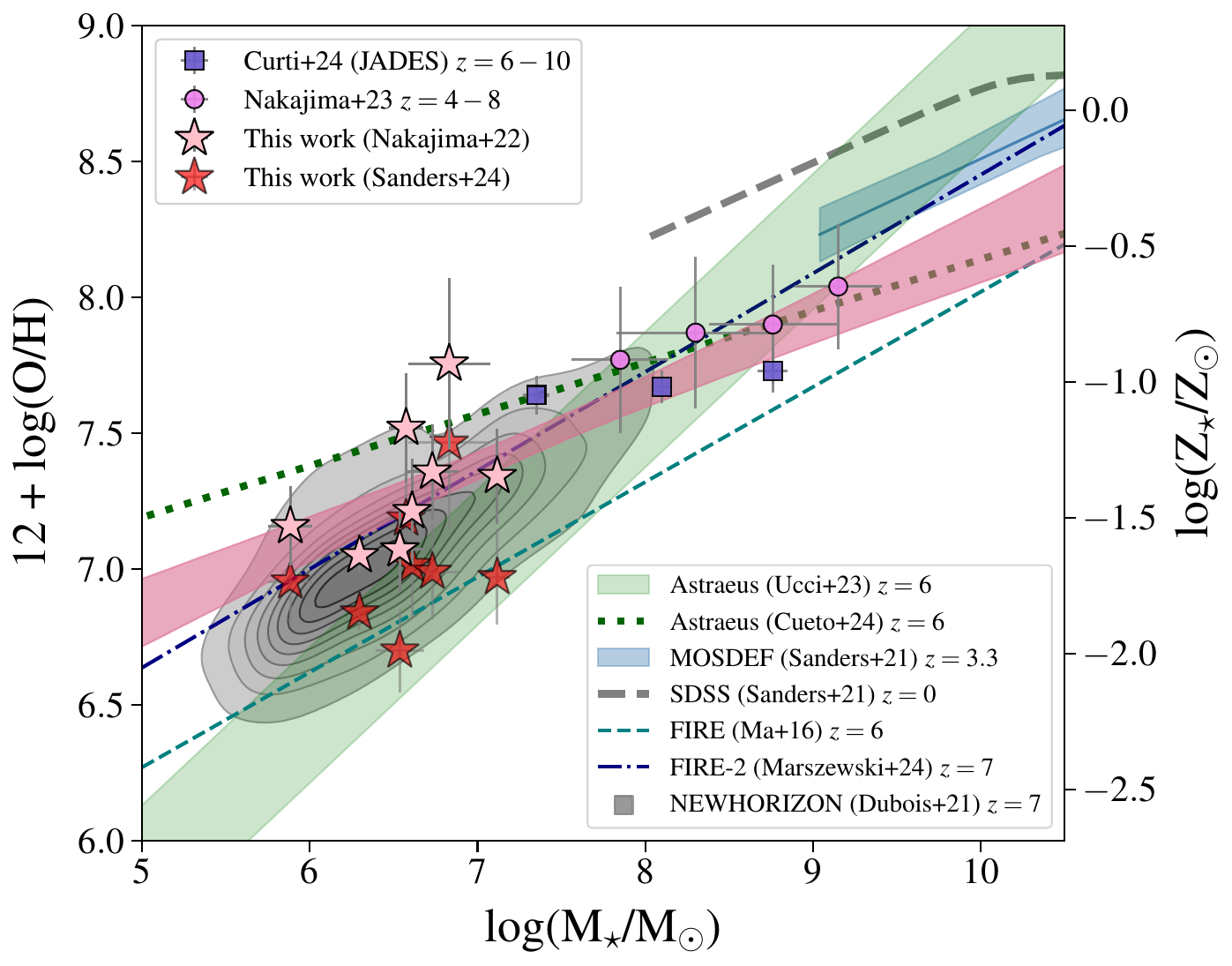} 
    \includegraphics[width=0.45\linewidth]{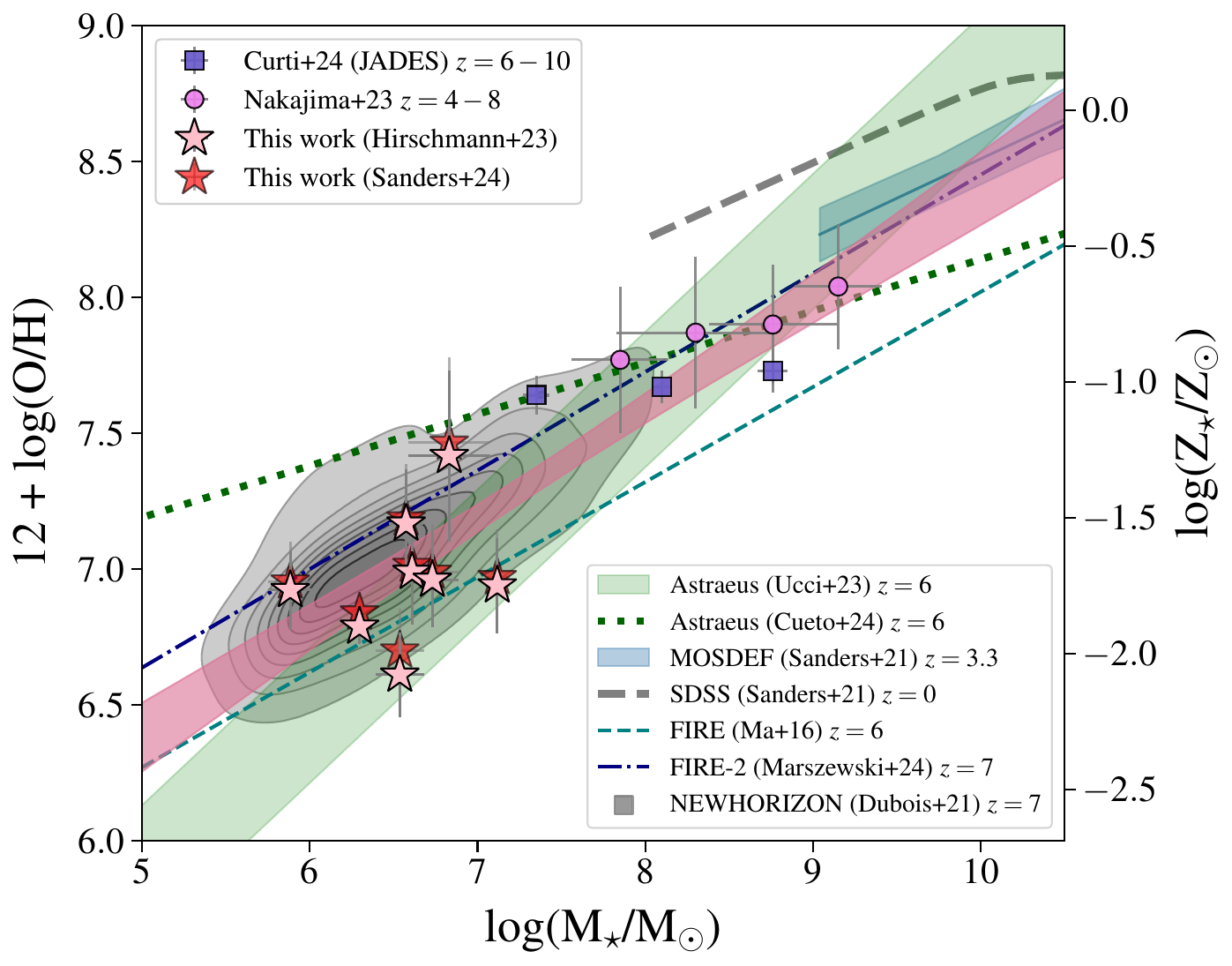}
    \caption{Identical to Figure \ref{fig:mz}. The mass-metallicity relation at $z = 6-8$ for the lowest-mass galaxies using a single power law fit. The pink stars represent the metallicities derived by using calibrations from \citet{Laseter24}, \citet{nakajima22}, and \citet{Hirschmann23} (panels from left to right, respectively), along concerning the \citet[][red stars]{sanders24}. The fitted MZR is shown in pink.}
    \label{fig:MZR_L_N}
\end{figure*}

In section \ref{sec:Metalicity}, we covered the mass-metallicity relation (MZR) for a combined sample of observed galaxies with the \jwst. Here we also estimate the MZR by using additional high-z calibrations to see to what extent our results are affected. We used the simple fit relation in the form:
\begin{equation*}
    12+\log(O/H)=m \times \log(M_{\star})+b
\end{equation*}
In Figure \ref{fig:MZR_L_N}, we display in each panel the resulting MZR for a given calibration (pink stars) compared to the fiducial calibration (red stars). The left top panel shows the MZR with adopting a single power law using \citet{sanders24} calibration, $m=0.39_{-0.02}^{+0.02}$, $ b = 4.52_{-0.17}^{+0.17}$. The slope of the relation is shallower, with $m=0.32^{+0.02}_{-0.02}, b=5.09^{+0.17}_{-0.17}$] when the metallicity estimate is based on \citet{Laseter24}. We see a similar result when adopting the \citet{nakajima22} calibration, which results in a best-fit relation of $m=0.27^{+0.02}_{-0.02}, b=5.48^{+0.17}_{-0.17}$. As was discussed above, those calibrations may overestimate the metallicities at the high redshifts. On the other hand, we obtain very similar results when adopting the \citet{Hirschmann23} prescription, with $m=0.40^{+0.02}_{-0.02}, b=4.36^{+0.17}_{-0.17}$.

\bibliography{references}{}
\bibliographystyle{aasjournal}

\end{document}